\documentclass[preprint,showpacs,preprintnumbers,amsmath,amssymb]{revtex4}
\usepackage{epsfig}
\usepackage{graphicx}
\usepackage{dcolumn}
\usepackage{bm}

\let\jnfont=\rm
\def\NPB#1,{{\jnfont Nucl.\ Phys.\ B }{\bf #1},}
\def\PLB#1,{{\jnfont Phys.\ Lett.\ B }{\bf #1},}
\def\EPJC#1,{{\jnfont Eur.\ Phys.\ Jour.\ C }{\bf #1},}
\def\PRD#1,{{\jnfont Phys.\ Rev.\ D }{\bf #1},}
\def\PRL#1,{{\jnfont Phys.\ Rev.\ Lett.\ }{\bf #1},}
\def\MPLA#1,{{\jnfont Mod.\ Phys.\ Lett.\ A }{\bf #1},}
\def\JPG#1,{{\jnfont J.\ Phys.\ G}{\bf #1},}
\def\CTP#1,{{\jnfont Commun.\ Theor.\ Phys.\ }{\bf #1},}
\def\ZPC#1,{{\jnfont Z.\ Phys.\ C }{\bf #1},}
\def\JHEP#1,{{\jnfont JHEP \ }{\bf #1},}
\def\q_slash{\not{\hbox{\kern-2.1pt $q$}}}
\def\p_slash{\not{\hbox{\kern-4.0pt $p$}}}
\def\k_slash{\not{\hbox{\kern-2.1pt $k$}}}

\begin{document}


\title{\ \\[10mm] Single top production associated with a neutral scalar
                  at LHC \\ in topcolor-assisted technicolor}

\author{Guoli Liu, Huanjun Zhang \vspace*{0.5cm} }
\affiliation{
Department of Physics, Henan Normal University, Xinxiang 453007,  China 
\vspace*{2cm}}

\begin{abstract}
The topcolor-assisted technicolor (TC2) model predicts a number of neutral scalars
like the top-pion ($\pi^0_t$) and the top-Higgs ($h^0_t$). These scalars have
flavor-changing neutral-current (FCNC) top quark couplings, among which the top-charm
transition couplings may be sizable. Such FCNC couplings induce single top productions
associated with a neutral scalar at the CERN Large Hadron Collider (LHC)
through the parton processes $cg \to t \pi_t^0$ and $cg \to t h_t^0$.
In this note we examine these productions and find their production rates
can exceed the $3\sigma$ sensitivity of the LHC in a large part of parameter space.
Since in the Standard Model and the minimal supersymmetric model
such rare productions have unobservably small production rates at the LHC,
these rare processes will serve as a good probe for the TC2 model.

\end{abstract}


\maketitle
As the heaviest fermion in the  Standard Model (SM), the top quark
may be a sensitive probe of new physics  \cite{review}.
So far there remain plenty of room for new physics in top quark sector due to
the small statistics of the top quark events at the Fermilab
Tevatron collider \cite{top-mass}.
Since the upcoming Large Hadron Collider (LHC) at CERN will produce
top quarks copiously and allow to scrutinize the top quark nature, the new physics
related to the top quark will be either uncovered or stringently constrained.

One of the properties of the top quark in the Standard Model (SM)
is its extremely small  flavor-changing neutral-current (FCNC) \cite{tcvh-sm}
interactions due to the GIM mechanism.
Thus, the observation of any FCNC top quark process would be a robust
evidence for new physics beyond the SM.
Actually,  the FCNC top quark interactions can be significantly enhanced
in some new physics models, such as the popular minimal supersymmetric
model (MSSM) \cite{tcv-mssm,pptc-mssm1,pptc-mssm2}
and the topcolor-assisted technicolor (TC2) model \cite{tc-tc2}.

The TC2 model predicts a number of neutral scalar bosons like the top-pions
and top-Higgs at the weak scale \cite{TC2}. These scalars
have  FCNC top couplings at tree-level, among which the top-charm
FCNC couplings are most significant.
Such anomalous FCNC couplings will induce single top productions
associated with a neutral scalar at the LHC
through the parton processes $cg \to t \pi_t^0$ and $cg \to t h_t^0$.
In this note we examine these productions and figure out if their rates
can exceed the $3\sigma$ sensitivity of the LHC.
Since in the SM and the MSSM such rare productions
have unobservably small production rates at the LHC,
these rare processes will serve as a probe for the TC2 model
if their TC2 predictions can be above the $3\sigma$ sensitivity.
\vspace*{0.5cm}

Before our calculations we recapitulate the basics of TC2 model.
The TC2 model \cite{TC2} combines technicolor
interaction with topcolor interaction, with the former being
responsible for electroweak symmetry breaking and the latter for
generating large top quark mass.
The top quark mass is generated from two sources, one is from
the extended technicolor (proportional to $\epsilon$) and the other
from the topcolor (proportional to $1-\epsilon$).
So the mass matrix of up-type quarks is composed of both
extended technicolor and topcolor contributions. The diagonalization
of this mass matrix will induce FCNC top quark interactions in the Yukawa
couplings which involve the composite scalars
respectively from topcolor and technicolor condensations.

The top-charm FCNC couplings with the top-pion
and top-Higgs can be written as \cite{tc-tc2}
\begin{eqnarray}
{\cal{L}}_{FCNC} = \frac{(1 - \epsilon ) m_{t}}{\sqrt{2}F_{t}}
     \frac{\sqrt{v^2-F_{t}^{2}}} {v}
      (i K_{UL}^{tt *} K_{UR}^{tc} \bar{t}_L c_{R} \pi_t^0
         + K_{UL}^{tt *} K_{UR}^{tc} \bar{t}_L c_{R} h_t^0 + h.c.) ,
\label{FCNH}
\end{eqnarray}
where the factor $\sqrt{v^2-F_t^2}/v$ ( $v \simeq 174$ GeV )
reflects the effect of the mixing between the top-pions and the
would-be Goldstone bosons \cite{9702265}.
$K_{UL}$ and $K_{UR}$ are the rotation matrices that
transform respectively the weak eigenstates of left-handed and right-handed
up-type quarks to their mass eigenstates, which can be parametrized as
\cite{tc-tc2}
\begin{equation}
K_{UL}^{tt}  \simeq 1, \hspace{5mm}
K_{UR}^{tt}\simeq \frac{m_t^\prime}{m_t} = 1-\epsilon,\hspace{5mm}
K_{UR}^{tc}\leq \sqrt{1-(K_{UR}^{tt})^2}
=\sqrt{2\epsilon-\epsilon^{2}}, \label{FCSI}
\end{equation}
with $m_t^\prime$ denoting the topcolor contribution to the top
quark mass. In Eq.(\ref{FCNH}) we neglected the mixing between up
quark and top quark.

\begin{figure}[bt]
\epsfig{file=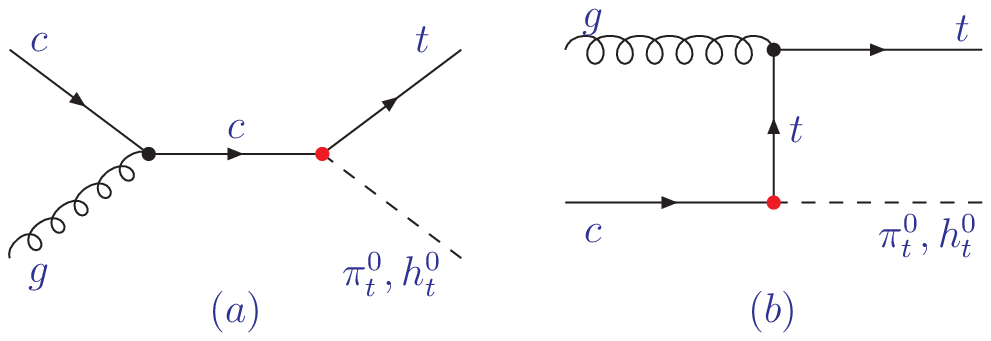,width=12cm} 
\vspace*{-0.5cm}
\caption{Feynman diagrams for
parton-level process $c g \to t \pi_t^0$ and  $c g \to t h_t^0$.}
\label{cgtpi}
\end{figure}

As shown in Fig. \ref{cgtpi}, these FCNC interactions in  Eq.(\ref{FCNH})
induce single top productions
associated with a top-pion or a top-Higgs at the LHC
through the parton processes $cg \to t \pi_t^0$ or
$cg \to t h_t^0$.
The amplitude for $cg\to t\pi_t^0$ is given by
\begin{equation}
{\cal M} = -\frac{1}{\sqrt{2}}g_s\frac{m_t}{F_t}\frac{\sqrt{v^2-F_t^2}}{v}
  K_{UR}^{tt*}K_{UR}^{tc} 
 \bar{u}_t  \left[\frac{1}{\hat{s}}
(\p_slash_c +\p_slash_g)\gamma^\mu
+\frac{1}{\hat{u}-m_t^2} \gamma^\mu(\p_slash_t-\p_slash_g+m_t) \right]P_R u_c ,
\label{amp-pi}
\end{equation}
where $P_R=(1+\gamma_{5})/2$, and
$p_{t,c,g}$ are the momentum of top quark, charm quark and gluon, respectively.

The hadronic cross section at the LHC is obtained by convoluting the
parton cross section with the parton distribution functions. In our
calculations we use CTEQ6L \cite{cteq} to generate the parton
distributions with the renormalization scale $\mu_R $ and the
factorization scale $\mu_F$ chosen to be $\mu_R = \mu_F = m_t + M_S$
($ M_S$ denotes top-pion mass or top-Higgs mass).

The parameters involved in our calculations are
the masses of the top-pions and top-Higgs,
the parameter $K_{UR}^{tc}$, the top-pion decay constant $F_t$
and the parameter $\epsilon$ which parametrizes the portion of
the extended-technicolor contribution to the top quark mass.
In our numerical calculations, we take $F_t=50$ GeV,
$\epsilon=0.1$, $K_{UL}^{tt}=1$, $K_{UR}^{tt}=0.9$ and retain
$K_{UR}^{tc} $ as a free parameter with a value less than
$\sqrt{2\epsilon-\epsilon^{2}}=0.43$.
The top quark mass is taken as $m_t=170.9$ GeV \cite{top-mass}.

For the masses of the neutral top-pion and top-Higgs, current 
constraints are rather weak. 
Theoretically the top-pion masses are model-dependent and
are usually of a few hundred GeV \cite{TC2}. The top-Higgs
mass, as analysed in \cite{tc-tc2}, has a lower bound of about
$2m_t$, which however is an approximate analysis and the mass below
$t\bar t$ threshold is also possible \cite{9809470}. 
Experimentally, the neutral top-pion mass can be constrained 
if we assume the degeneracy of neutral and charged  top-pion masses 
(the mass splitting between the neutral top-pion and the charged
top-pion comes only from the electroweak interactions and thus
should be small).  The charged top-pion mass is constrained 
from the absence of $t \to \pi_t^+b$, which gives 
a lower bound of 165 GeV \cite{t-bpion}, and also from
$R_b$ data, which yields a lower bound of about 250 GeV \cite{burdman}.
In our numerical results we will
show the dependence on the masses of neutral top-pion and top-Higgs.

In the following we present some results for the hadronic production
cross section via  $c g \to t  \pi_t^0$. These results are also applicable
to the production through $c g \to t  h_t^0$, with the top-pion mass replaced by
the top-Higgs mass.

\begin{figure}[bt]
\epsfig{file=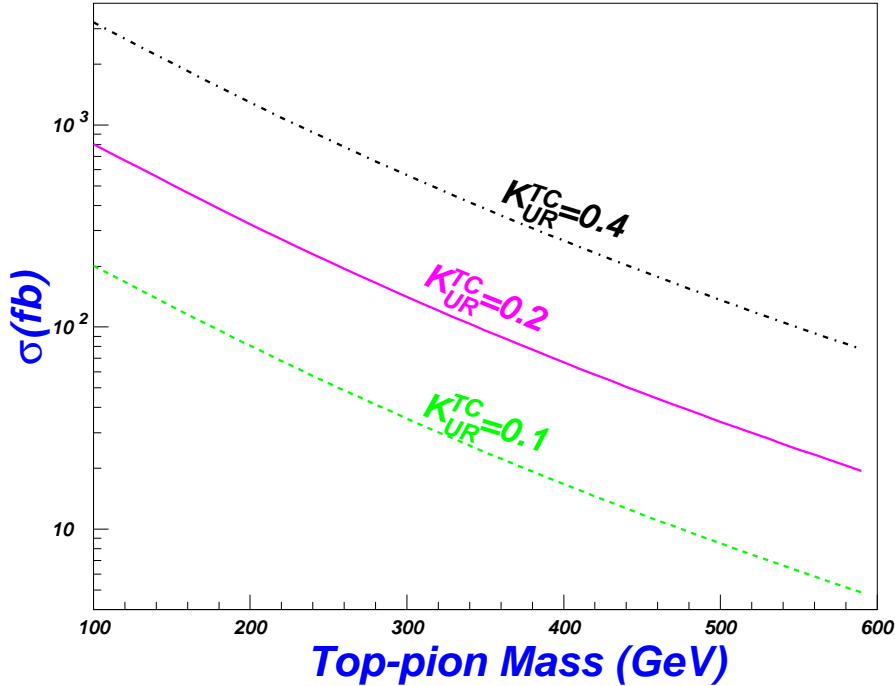,width=12cm}
\vspace*{-0.5cm}
\caption{Hadronic cross section for the production via $c g \to t  \pi_t^0$
at the LHC versus top-pion mass.}
\label{cgtpi1}
\end{figure}

Fig.\ref{cgtpi1} shows that the hadronic cross section
versus top-pion mass for different values of $K_{UR}^{tc}$.
We see that the cross section increases with the increasing
$K_{UR}^{tc}$ since the cross section
is simply proportional to $(K_{UR}^{tc})^2$ as shown in
Eq.(\ref{amp-pi}).
As the top-pion mass increases, the cross section decreases.
The cross section is about several hundreds fb in most of the
parameter space.

Due to the large QCD backgrounds at the LHC,
for the productions of $PP\to t\phi^0+X$ ($\phi^0$ is a neutral scalar
and can be top-pion or top-Higgs) we search for the final states
from the subsequent decays $t\to Wb \to \ell \nu b$ (
$ \ell=e,\mu$) and $\phi^0\to b\bar b$. 
So the main SM background
is the production of $t \bar t$ and $Wb \bar b  jj$, where one light jet
is mis-identified as a b-jet while the other light jet is not
detected if it goes along the pipeline or its transverse momentum is
too small.

The observability of the signal at the LHC has been investigated
in the effective Lagrangian  approach  \cite{Aguilar-Saavedra:2004wm}.
Assuming a luminosity of $100$ fb$^{-1}$, we know from Table 2, Table 4
and Eq.(4) in \cite{Aguilar-Saavedra:2004wm} that the $3\sigma$ sensitivity
for the production of $PP\to t\phi^0+X$ is about 200 fb.
Although this sensitivity is based on the effective Lagrangian
approach and may be not perfectly applicable to a specified model,
we can take them as a rough criteria to estimate the observability of
these channels.

\begin{figure}[hbt]
\epsfig{file=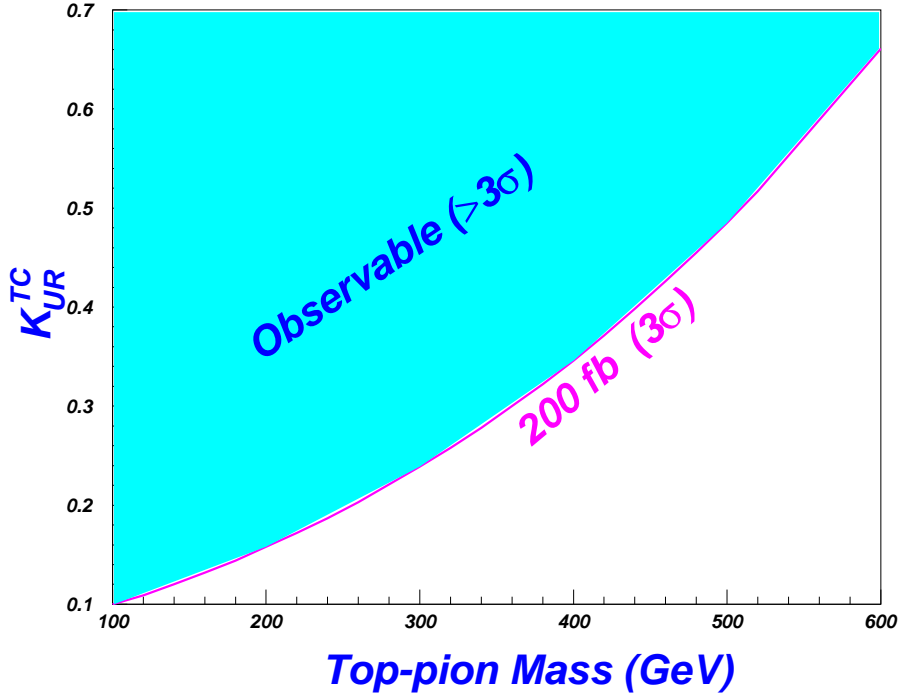,width=12cm} 
\vspace*{-0.5cm}
\caption{The contour of
$3\sigma$ sensitivity (200 fb) for the cross section
 of the production via $c g \to t  \pi_t^0$
at the LHC in the plane
of $K_{UR}^{tc}$ versus top-pion mass.} 
\label{contour}
\end{figure}

To show the observability of the production of $PP\to t\pi^0_t+X$,
we plot in Fig.\ref{contour} the contour of the cross section of
the $3\sigma$ sensitivity (200 fb) in the plane of $K_{UR}^{tc}$ versus
$m_{\pi^0_t}$. We see that in a large part of the parameter space
the cross section can exceed the $3\sigma$ sensitivity.
As we mentioned earlier, the production $PP\to th+X$ is unobservably small
in the SM due to the extremely suppressed $h t\bar c$ coupling. In the
MSSM the $PP\to th+X$ has a rate lower than 10 fb \cite{pptc-mssm2} and
thus also inaccessible at the LHC.
So these rare processes will serve as a good probe for the TC2 model.

In conclusion,
we examined the single top productions associated with a neutral scalor
(top-pion or top-Higgs) at the LHC in topcolor-assisted technicolor
model. We found that their production rates
can exceed the $3\sigma$ sensitivity of the LHC in a large part of
parameter space.
Since in the Standard Model and the minimal supersymmetric model
such rare productions have unobservably small production rates at the LHC,
these rare processes will serve as a good probe for the
topcolor-assisted technicolor model.

\begingroup\raggedright
\endgroup

\end{document}